\documentclass[11pt]{article}
\usepackage{graphicx}
\usepackage{epsfig}
\usepackage{times}
\usepackage{fancyheadings}
\usepackage{lscape}
\usepackage{rotating}
\usepackage{longtable}
\usepackage{multirow}
\usepackage{xspace}
\usepackage{subfigure}

\newcommand{\BABARPubYear}    {06}

\newcommand{\BABARConfNumber} {028}
\newcommand{\SLACPubNumber} {12018}

\input figuretex/babarsym

\def\Vub {\ensuremath{V_{ub}}}

\def\btn {\ensuremath{B^{+} \to \tau^{+} \nu_{\tau}}}
\def\btodlnux {\ensuremath{\Bub \to \Dz \ell^{-} \bar{\nu}_{\ell} X}}

\def\eextra {\ensuremath{E_{\mathrm{extra}}}}

\def\nulb  {\ensuremath{\bar{\nu}_{\ell}}}
\def\nueb  {\ensuremath{\bar{\nu}_{e}}}

\def\nutb  {\ensuremath{\bar{\nu}_{\tau}}}

\def\btodszlnu {\ensuremath {\B^- \to D^{*0} \ell^- \nulb}}

\def\btodlnu {\ensuremath {\B^- \to D^{0} \ell^- \nulb}}

\def\dlnux {\ensuremath {D^{0} \ell^- \nulb X}}
\def\tautoenunu {\ensuremath {\tau^+ \to e^+ \nu_e \nutb}}
\def\tautoe {\ensuremath {\tau^+ \to e^+ \nu_e \nutb}}
\def\enunu {\ensuremath {e^+ \nu_e \nutb}}
\def\tautomununu {\ensuremath {\tau^+ \to \mu^+ \nu_{\mu} \nutb}}
\def\tautomu {\ensuremath {\tau^+ \to \mu^+ \nu_{\mu} \nutb}}
\def\mununu {\ensuremath {\mu^+ \nu_{\mu} \nutb}}
\def\tautopinu {\ensuremath {\tau^+ \to \pi^+ \nutb}}
\def\tautopi {\ensuremath {\tau^+ \to \pi^+ \nutb}}
\def\pinu {\ensuremath {\pi^+ \nutb}} 
\def\tautopipiznu {\ensuremath {\tau^+ \to \pi^+ \pi^{0} \nutb}}
\def\tautopipiz {\ensuremath {\tau^+ \to \pi^+ \pi^{0} \nutb}}
\def\tautorho {\ensuremath {\tau^+ \to \pi^+ \pi^{0} \nutb}}
\def\pipiznu {\ensuremath {\pi^+ \pi^{0} \nutb}}

\def\onlumi    {\ensuremath { 288  \invfb\ }}
\def\offlumi   {\ensuremath { 27.5 \invfb\  }}

\def\nBB   {\ensuremath {320 \times 10^{6}} }

\def\etal {{\it et al.}}

\long\def\inst#1{\par\nobreak\kern 4pt\nobreak
    {\it #1}\par\vskip 10pt plus 3pt minus 3pt}

\begin{document}
{\pagestyle{empty}

\begin{flushright}
\babar-CONF-\BABARPubYear/\BABARConfNumber \\
SLAC-PUB-\SLACPubNumber \\
\end{flushright}

\par\vskip 5cm

\begin{center}
\Large \bf A Search for \boldmath{$B^+\to\tau^+\nu$} Recoiling Against 
\boldmath{\btodlnux}.
\end{center}
\bigskip

\begin{center}
\large The \babar\ Collaboration\\
\mbox{ }\\
\today
\end{center}
\bigskip \bigskip

\begin{center}
\large \bf Abstract
\end{center}

\noindent We present a search for the decay \btn\ using 288 \invfb of data 
collected at the $\Y4S$ resonance with the \babar\ detector 
at the SLAC PEP-II $B$-Factory. A sample of events with one reconstructed 
semileptonic $B$ decay (\btodlnux) is selected, and
in the recoil a search for $\btn$ signal is performed. The $\tau$ is
identified in the following channels: $\tautoenunu$, $\tautomununu$,
$\tautopinu$ and $\tautopipiznu$. 
We measure a branching fraction of 
$\mathcal{B}(\btn)=(0.88^{+0.68}_{-0.67}(\mbox{stat.}) \pm 0.11 (\mbox{syst.})) \times 10^{-4}$
and extract an upper limit on the branching fraction, at the
90\% confidence level, of $\mathcal{B}(\btn) < 1.8 \times 10^{-4}$.
We calculate the product of the $B$ meson decay constant and $|\Vub|$ to be
$f_{B}\cdot|\Vub| = (7.0^{+2.3}_{-3.6}(\mbox{stat.})^{+0.4}_{-0.5}(\mbox{syst.}))\times10^{-4}$~GeV.

\vfill
\begin{center}

Submitted to the 33$^{\rm rd}$ International Conference on High-Energy Physics, ICHEP 06,\\
26 July---2 August 2006, Moscow, Russia.

\end{center}

\vspace{1.0cm}
\begin{center}
{\em Stanford Linear Accelerator Center, Stanford University, 
Stanford, CA 94309} \\ \vspace{0.1cm}\hrule\vspace{0.1cm}
Work supported in part by Department of Energy contract DE-AC03-76SF00515.
\end{center}

\newpage
} 

\begin{center}
\small

The \babar\ Collaboration,
\bigskip

%
{B.~Aubert,}
{R.~Barate,}
{M.~Bona,}
{D.~Boutigny,}
{F.~Couderc,}
{Y.~Karyotakis,}
{J.~P.~Lees,}
{V.~Poireau,}
{V.~Tisserand,}
{A.~Zghiche}
\inst{Laboratoire de Physique des Particules, IN2P3/CNRS et Universit\'e de Savoie,
 F-74941 Annecy-Le-Vieux, France }
% ...

\end{center}\newpage

\section{INTRODUCTION}
\label{sec:Introduction}

\begin{figure}[htb]
\begin{center}
\includegraphics[width=0.45\textwidth]{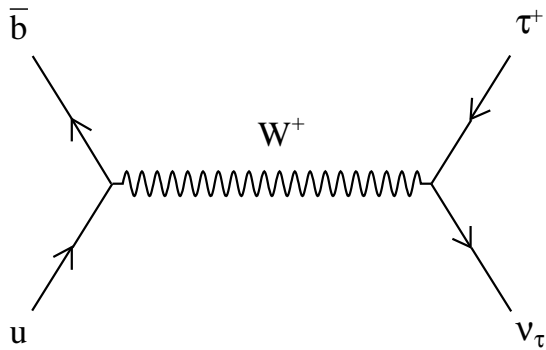}
\end{center}
\caption{\label{fig:feynman_diagram}%
The purely leptonic $B$ decay $\Bu \to \tau^{+} \nu_{\tau}$  
proceeding via quark annihilation into a $W^{+}$ boson.}
\end{figure}

In the Standard Model (SM), the purely leptonic decay \btn\ 
\footnote{Charge-conjugate modes are implied throughout this paper. The signal $B$ will always be denoted as a \Bu\ decay while the semi-leptonic $B$ will be denoted as a \Bub\ to avoid confusion.}
proceeds via quark annihilation 
into a $W^{+}$ boson (Fig. \ref{fig:feynman_diagram}).
Its amplitude is thus proportional to the product of the 
$B$-decay constant $f_B$ and the quark-mixing-matrix 
element $V_{ub}$. The branching fraction is given by:
\begin{equation}
\label{eqn:br}
\mathcal{B}(B^{+} \rightarrow {\taup} \nu)= 
\frac{G_{F}^{2} m^{}_{B}  m_{\tau}^{2}}{8\pi}
\left[1 - \frac{m_{\tau}^{2}}{m_{B}^{2}}\right]^{2} 
\tau_{\Bu} f_{B}^{2} |V_{ub}|^{2},\label{eq:brsm} 
\end{equation}
where we have set $\hbar = c = 1$, %
$G_F$ is the Fermi constant, 
$V_{ub}$ is a quark mixing matrix element~\cite{ref:c,ref:km}, 
$f_{B}$ is the $\Bu$ meson decay constant which describes the
overlap of the quark wave-functions inside the meson,
$\tau_{\Bu}$ is the $\Bu$ lifetime, and
$m^{}_{B}$ and $m_{\tau}$ are the $\Bu$ meson and $\tau$ masses.
This expression is entirely analogous to that for pion decay.
Physics beyond the SM, such as a two-Higgs doublet models,
could enhance or suppress the $\mathcal{B}(\btn)$ through the
introduction of a charged Higgs boson~\cite{ref:higgs}.

Current theoretical values for $f_B$ 
(obtained from lattice QCD calculations)~\cite{ref:fb} have  
large uncertainties, and purely leptonic decays of the $\Bu$ meson
may be the only clean experimental method of measuring
$f_B$ precisely. Given measurements of $|V_{ub}|$ from semileptonic
$B \to\u\ell\nu$ decays, $f_{B}$ could be extracted
from the measurement of the \btn\ branching fraction. In addition,
by combining the branching fraction measurement with results
from $B$ mixing, the ratio $|V_{ub}|/|V_{td}|$ can be extracted
from $\mathcal{B}(\btn)/\Delta m$, where $\Delta m$ is the
mass difference between the heavy and light neutral $B$ meson states.

The decay amplitude is proportional to the lepton mass and
as such decay to the lighter leptons is suppressed. This mode
is the most promising for discovery of leptonic $B$ decays.
However, experimental challenges 
such as the large missing momentum from several neutrinos make 
the signature for \btn\ less distinctive than for other leptonic modes.

The SM estimate of this branching fraction
is $(1.59 \pm 0.40)\times 10^{-4}$, 
using $|\Vub| = (4.39 \pm 0.33)\times 10^{-3}$~\cite{ref:vub} and 
$f_{B}~=~0.216\pm 0.022$~GeV~\cite{ref:fb} in Eq.~\ref{eq:brsm}.

In a previously published analysis of a smaller sample of 
$223 \times 10^6$ $\FourS$ decays
the \babar\ collaboration set an upper limit of:
\begin{eqnarray}
\mathcal{B}(\btn) & < & 2.6 \times 10^{-4} \, \textrm{at the 90\% CL. }~\cite{ref:babar_prd_btn}
\end{eqnarray}
The Belle collaboration reported evidence of a signal in this channel 
recently~\cite{ref:belle}; a central value of \\
$(1.06^{+0.34}_{-0.28}\rm{(stat.)}^{+0.18}_{-0.16}\rm{(syst.)})\times10^{-4}$ was extracted.
The analysis presented herein is a detailed update of the previous \babar\ search.

\section{THE \babar\ DETECTOR AND DATASET}
\label{sec:babar}
The data used in this analysis were collected with the \babar\ detector
at the \pep2\ storage ring. 
The sample corresponds to an integrated
luminosity of \onlumi \xspace at the \FourS\ resonance (on-resonance) 
and \offlumi \xspace taken $40\mev$ below $B\bar{B}$ threshold 
(off-resonance). The on-resonance sample consists of
about $\nBB$  $\FourS$ decays (\BB\ pairs). The collider is operated with asymmetric
beam energies, producing a boost of $\beta\gamma \approx 0.56$ 
of the \FourS along the collision axis.

The \babar\ detector is optimized for asymmetric energy collisions at a
center-of-mass (CM) energy corresponding to the \FourS\ resonance.
The detector is described in detail in Ref.~\cite{ref:babar}. 
The components used in this analysis are the tracking system
composed of a five-layer silicon vertex detector and a 40-layer drift chamber (DCH),
the Cherenkov detector (DIRC) for charged $\pi$--$K$ discrimination, the CsI calorimeter
(EMC) for photon and electron identification, and the 18-layer flux return (IFR) located 
outside of the 1.5T solenoidal coil and instrumented with resistive plate chambers for muon
and neutral hadron identification. For the most recent 51 \invfb of data, a portion of the muon 
system has been upgraded to limited streamer tubes (LST)~\cite{ref:lst}.
We separate the treatment of the data to account for varying accelerator and detector conditions.
``Runs~1--3'' corresponds to the first 111.9\invfb, ``Run~4'' the following 99.7\invfb and ``Run~5''
the subsequent 76.8\invfb.

A GEANT4-based \cite{ref:geant4} Monte Carlo (MC) 
simulation is used to model the signal
efficiency and the physics backgrounds. Simulation samples
equivalent to approximately three times the accumulated data  were
used to model \BB\ events, and samples equivalent to approximately
1.5 times the accumulated data were used to model continuum events where
$\epem \to$ \uubar, \ddbar, \ssbar, \ccbar and \tautau. 
A large sample of signal events is simulated, where
a $B^+$ meson decays to $\tau^+\nu_{\tau}$ and a $B^-$  
meson decays to an acceptable $B$ mode. Beam related background and detector 
noise from data are overlayed on the simulated events.

\section{ANALYSIS METHOD}
\label{sec:Analysis}

Due to the presence of multiple neutrinos, the \btn \xspace decay mode
lacks the kinematic constraints which are usually exploited in $B$ decay 
searches in order to reject both continuum and $B\overline{B}$ backgrounds.
The strategy adopted for this analysis is to reconstruct exclusively 
the decay of one of the $B$ mesons in the event, referred to as ``tag'' $B$. 
The remaining particle(s) in the event, referred to as the 
``signal side'', are then compared with the signature
expected for \btn. In order to avoid experimenter bias, the 
signal region in data is not examined (``blinded'') until the final yield
extraction is performed.

The tag $B$ is reconstructed in the set of semileptonic $B$ decay modes 
\btodlnux, where $\ell$ is $e$ or $\mu$ and $X$ can be either nothing or a 
transition particle from a higher mass charm state decay which we do not attempt 
to reconstruct (although those tags consistent with neutral $B$ decays are vetoed).
The $\Dz$ is reconstructed in four decay modes:
$K^{-}\pi^{+}$, $K^{-}\pi^{+}\pi^{-}\pi^{+}$, $K^{-}\pi^{+}\pi^{0}$, and
$K_{s}^{0}\pi^{+}\pi^{-}$. The $K_{s}^{0}$ is reconstructed only in the
mode $K_{s}^{0} \rightarrow \pi^{+}\pi^{-}$.
These cases where the low momentum transition daughter
of $D^{*0}$ decays need not be reconstructed and the final state 
$B\to\Dz\ell\nu X$ as observed provides a higher efficiency but
somewhat lower purity than the exclusive reconstruction method of
$\btodszlnu$.
The choice of reconstructing the tag $B$ as $\btodlnux$ was optimized 
by maximizing $s/\sqrt{s+b}$ where $s =$~signal and $b =$~background where
a branching fraction for $\btn$ of $1\times10^{-4}$ is assumed.

The \btn \xspace signal is searched for in
both leptonic and hadronic $\tau$ decay modes:
$\tautoenunu$, $\tautomununu$, $\tautopinu$ and $\tautopipiznu$.
The branching fractions of the above $\tau$ decay 
modes are listed in Table \ref{tab:TauDecayModes}. 

\begin{table}[h]
\caption{\label{tab:TauDecayModes} Branching fractions for the $\tau$ decay modes used in the \btn\ search~\cite{ref:pdg2004}.}
\begin{center}
\begin{tabular}{|l|c|}
\hline
Decay Mode   &   Branching Fraction (\%)  \\
\hline\hline
$\tautoenunu$      & 17.84 $\pm$ 0.06 \\ \hline
$\tautomununu$     & 17.36 $\pm$ 0.06 \\ \hline
$\tautopinu$       & 11.06 $\pm$ 0.11 \\ \hline
$\tautopipiznu$    & 25.42 $\pm$ 0.14 \\ \hline
\end{tabular}
\end{center}
\end{table}

\subsection{Tag \boldmath{B} Reconstruction}
\label{sec:TagReco}

The tag $B$ reconstruction proceeds as follows. First we reconstruct the 
$\Dz$ candidates in the aforementioned four decay modes using reconstructed tracks
and photons where a $\piz$ is included. The tracks
are required to meet particle identification criteria consistent
with the particle hypothesis, and are required to converge at a common vertex.
The $\piz$ candidate is required to have invariant mass between 
0.115--0.150 \gev/$c^2$ and its daughter photon candidates must 
have a minimum energy of 30 \mev.
The mass of the reconstructed $\Dz$ candidates in 
$K^{-}\pi^{+}$, $K^{-}\pi^{+}\pi^{-}\pi^{+}$, and $K_{s}^{0}\pi^{+}\pi^{-}$
modes are required to be within 20 \mev/$c^2$ of the nominal mass 
\cite{ref:pdg2004}.
In the $K^{-}\pi^{+}\pi^{0}$ decay mode 
the mass is required to be within 35 \mev/$c^2$ of the nominal mass
\cite{ref:pdg2004}.

Finally $\Dz\ell$ candidates are reconstructed by combining the
$\Dz$ with an identified electron 
or muon with momentum above 0.8 \gev/$c$ in the CM frame. 
The $D^{0}$ and $\ell$ candidates are required to meet at a common vertex.
An additional kinematic constraint is imposed on the reconstructed 
$\Dz\ell$ candidates: 
assuming that the massless neutrino is the only missing particle, we 
calculate the cosine of the angle between the $\Dz\ell$ candidate
and the $B$ meson,
\begin{equation}
\cos\theta_{B-D^{0}\ell} = \frac{2 E_{B} E_{D^{0}\ell} - m_{B}^{2} - m_{D^{0}
\ell}^{2}}{2|\vec{p}_{B}||\vec{p}_{D^{0}\ell}|}.
\end{equation}
Here ($E_{D^{0}\ell}$, $\vec{p}_{D^{0}\ell}$) and
($E_{B}$, $\vec{p}_{B}$) are the 
four-momenta in the CM frame, and $m_{D^{0}\ell}$ and $m_{B}$ 
are the masses of the $D^{0}\ell$ candidate and $B$ meson, respectively. 
$E_{B}$ and the magnitude of $\vec{p}_{B}$ are calculated 
from the beam energy: $E_{B} = E_{\rm{CM}}/2$ and 
$ | \vec{p}_{B} | = \sqrt{E_{B}^{2} - m_{B}^{2} }$, where 
$E_{B}$ is the $B$ meson energy in the CM frame.
Correctly reconstructed candidates
populate the range [$-1,1$], whereas combinatorial backgrounds
can take unphysical values outside this range. 
We retain events in the interval 
$-2.0 < \cos\theta_{B-D^{0}\ell} < 1.1$, where the upper bound takes 
into account the detector resolution and the loosened lower bound
accepts those events where a soft transition particle from a higher mass
charm state is missing.

If more than one suitable $\Dz\ell$ candidate is 
reconstructed in an event, the best candidate is taken to be the
one with the largest vertex probability. 
The sum of the charges of all the particles in the event (net charge) 
must be equal to zero.

At this stage of the selection, the observed yield in data and
the predicted yield in the MC simulation agree to 
within approximately 3\%.
This discrepancy is corrected by scaling the yield and efficiency 
obtained from MC simulation.
By multiplying the relevant branching fractions and reconstruction 
efficiencies, from signal MC simulation, $B$ tagging efficiencies are 
extracted.
Scale factors of 1.05, 1.00 and 0.97 are used to correct these efficiencies
for Runs 1--3, Run~4 and Run~5 respectively.
The systematic error associated with this correction  
is described in Sec. \ref{sec:Systematics}. 
The corrected
tag reconstruction efficiency in the signal MC simulation is 
(7.61 $\pm$ 0.05)$\times 10^{-3}$ for Runs 1--3,
(6.31 $\pm$ 0.05)$\times 10^{-3}$ for Run 4 and
(5.87 $\pm$ 0.06)$\times 10^{-3}$ for Run 5 where the errors are statistical only.

\subsection{Selection of \boldmath{\btn} \xspace signal candidates}
\label{sec:SigSelection}

After the tag $B$ reconstruction, in the signal side
the $\tau$ from the $\btn$ decay is identified in one of the following modes:
$\tautoenunu$, $\tautomununu$, $\tautopinu$ or $\tautopipiznu$.
We select events with one signal-side track
which must satisfy the following selection criteria:
it must have at least 12 DCH hits, its momentum transverse to the 
beam axis, $p_{\rm{T}}$, is greater than 0.1 \gev/c, and
its point of closest approach to the interaction point is 
less than 5.0~\cm\ along the beam axis and less than 1.5~\cm\ transverse 
to the beam axis.
The invariant mass of a signal-side $\piz$ candidate
must be between 0.115--0.150 \gev/$c^2$,
the shower shape of the daughter photon candidates must be consistent with 
an electromagnetic shower shape and the photons
must have a minimum energy of 50 \mev in the CM frame. 

The different signal tau decay modes are distinguished by their 
selection criteria.
The $\tautoenunu$, $\tautomununu$, $\tautopinu$ and $\tautopipiznu$
signal modes, all of which contain one charged track, are separated by
particle identification.
Both the $\tautopinu$ and the $\tautopipiznu$ modes
contain a pion signal track and are characterized by the 
number of signal-side $\piz$ mesons.

\begin{itemize}

\item{Particle identification:}

\begin{itemize} 
\item For the $\tautoenunu$ selection the track must be 
identified as an electron and not identified as a muon.

\item For the $\tautomununu$ selection the track must be 
identified as a muon and not identified as an electron.

\item For the $\tautopinu$ selection we require that
the track is not identified as an electron or a muon.

\item For the $\tautopipiznu$ selection we require that
the track is not identified as an electron or a muon or a kaon.


\end{itemize} 

\item{Signal-side $\piz$ multiplicity:} 

\begin{itemize}

\item For the $\tautopinu$ selection we require the event to contain 
no signal-side $\piz$. 

\item For the $\tautopipiznu$ selection we require that the 
event contains at least one signal-side $\piz$.

\end{itemize}
\end{itemize}

\noindent Background consists primarily of $B^{+}B^{-}$ events in which the tag
$B$ meson has been correctly reconstructed and the recoil side contains
one signal candidate track and additional particles which are not 
reconstructed by the tracking detectors or calorimeters. Typically these
events contain  $K_{L}^{0}$ candidates and/or neutrinos, and frequently
also additional charged or neutral particles which pass outside of the 
tracking and calorimeter acceptance. Background events also contain
$B^{0}\bar{B}^{0}$ events. The continuum background contributes to 
hadronic $\tau$ decay modes. In addition some excess events in data,
most likely from two-photon and QED processes which are not modeled in the MC 
simulation, are also seen. These backgrounds have a distinctive event shape
and are suppressed by the
following constraints on the kinematics of the $\btn$ candidates.

\begin{itemize}

\item{Missing mass:} The missing mass is calculated as follows.
\begin{equation}
M_{\rm{miss}} = \sqrt{ (E_{\FourS}-E_{\rm{vis}})^2 - ( \vec{p}_{\FourS} - \vec{p}_{\rm{vis}} )^2 }.
\end{equation}
Here ($E_{\FourS}$, $\vec{p}_{\FourS}$) is the four-momentum of the $\FourS$,
known from the beam energies. The quantities $E_{\rm{vis}}$ and $\vec{p}_{\rm{vis}}$ are the 
total visible energy and momentum of the event which are calculated by adding the 
energy and momenta, respectively, of all the reconstructed 
charged tracks and photons in the event.

\begin{itemize} 

\item For the $\tautoenunu$ selection events with 
missing mass between 4.6 and 6.7 \gev/$c^2$ are selected.

\item For the $\tautomununu$ selection events with 
missing mass between 3.2 and 6.1 \gev/$c^2$ are selected.

\item For the $\tautopinu$ selection 
the missing mass is required to be greater than 1.6 \gev/$c^2$.

\item For the $\tautopipiznu$ selection 
the missing mass is required to be less than 4.6 \gev/$c^2$.


\end{itemize}

\item{Maximum CM momentum of the $\tau$ daughter:} 

The following maximum CM momentum requirements are applied to 
the $\tau$ daughter particles.

\begin{itemize}

\item The electron candidate from the $\tautoenunu$ decay must have a CM momentum of less than 1.5 \gev/c. 
The CM momentum requirement is not applied to the 
$\tautomununu$ selection because
the momentum spectrum of the muon from $\tau$ decays peaks below 1 \gev/c and 
the particle identification efficiency for low momentum muons is lower than 
that for low momentum electrons. Therefore, applying the maximum momentum 
cut reduces the selection efficiency of the $\tautomununu$ mode 
significantly.

\item For the two hadronic $\tau$ decay modes, 
the CM momentum of the $\pi$ from $\tautopinu$ must be greater than 1.6 \gev/c. 
The $\pi \piz$ combination from 
$\tautopipiznu$ must have CM momentum greater than 1.7 \gev/c.

\end{itemize}

\item{Continuum Rejection using the $R_{\tau\tau}$ variable:}

An effective way to remove $\ep\en\to\taup\taum$
background is to place a cut in a plane defined by two variables:
the cosine of the angle between the signal candidate and the tag $B$'s thrust
vector (in the CM frame), and the minimum invariant mass constructable 
from any three tracks in an event (regardless of whether they are already used
in a tag or signal candidates). 
For the background, the cosine of the thrust
angle peaks at $-1$ and 1, while the minimum invariant mass peaks below $1.5\gevcc$.
We transformed this 2-D variable into a 1-D variable using the following empirically derived equation
\begin{equation}
R_{\tau\tau} \equiv \sqrt{(3.7-|\cos(\theta_{\vec{T}_{D\ell},\rm{signal}})|)^{2} + (M^{\rm{min}}_{3}-0.75)^{2}},
\end{equation}
where $M^{\rm{min}}_3$ is the minimum invariant mass of any three changed tracks and
$\theta_{\vec{T}_{D\ell},\rm{signal}}$ is the angle between the thrust  
axes of the reconstructed $D\ell$ and the signal candidates.
Because other continuuum backgrounds also peak in the cosine of the thrust
angle, this variable is good at rejecting other similar categories of
non-\bbbar\ background. 
The selection criteria imposed on this quantity are:
%

\begin{itemize}
\item For $\tautoe$: $2.78 < R_{\tau\tau} < 4.0$
\item For $\tautomu$: $R_{\tau\tau} > 2.74$ 
\item For $\tautopi$: $R_{\tau\tau} > 2.84$
\item For $\tautorho$: $R_{\tau\tau} > 2.94$ 
\end{itemize}

The $\tautopipiznu$ decay proceeds via an intermediate 
resonance. For this mode further background rejection can be 
achieved by applying the following requirements on the intermediate meson.

\begin{itemize}

\item{$\rho^{+}$ selection:}

The signal-side track is combined with a signal-side $\piz$ to form
the $\rho^{+}$ candidate.
In events with more than one signal-side $\piz$, the candidate with 
invariant mass closest to the nominal $\piz$ mass \cite{ref:pdg2004}
is chosen. The invariant mass of the 
reconstructed $\rho^{+}$ is required to be within
0.64--0.86 \gev/$c^2$. A quantity similar to $\cos \theta_{B-D^{0} \ell}$,
which is defined in section \ref{sec:TagReco}, can be
reconstructed for $\tau \to \rho \nu$ as follows:
\begin{equation}
\cos\theta_{\tau-\rho} = \frac{2 E_{\tau} E_{\rho} - m_{\tau}^{2} - m_{\rho}^{2}}{2|\vec{p}_{\tau}||\vec{p}_{\rho}|},
\end{equation}
where ($E_{\tau}$, $\vec{p}_{\tau}$) and
($E_{\rho}$, $\vec{p}_{\rho}$) are the
four-momenta in the CM frame, $m_{\tau}$ and $m_{\rho}$
are the masses of the $\tau$ and $\rho$ candidate, respectively.
The quantities $|\vec{p}_{\tau}|$ and $E_{\tau}$ are calculated
assuming the $\tau$ is from the $\btn$ decay, and 
the $B^{+}$ is almost at rest in the CM frame.
We accept candidates with $\cos\theta_{\tau-\rho} > 0.87$.

%
%

\end{itemize}


\item{$E_{\rm{extra}}$ requirement:}

The most powerful variable for separating signal
and background is the remaining energy ($E_{\rm{extra}}$),
calculated by adding the CM energy of the neutral clusters and charged tracks that are not
associated with either the tag $B$ or the signal. The photon candidates contributing
to the $E_{\rm{extra}}$ variable have minimum cluster energies of 20 \mev in the CM frame.
For signal events
the neutral clusters contributing to $E_{\rm{extra}}$ arise 
predominantly from processes such as  beam-background, hadronic split-offs and
Bremsstrahlung. Signal events tend to peak at
low $E_{\rm{extra}}$ values whereas background events, which contain
additional sources of neutral clusters, are distributed
towards higher $E_{\rm{extra}}$ values. 
The most signal sensitive region is optimized for each mode and 
is blinded in on-resonance data until the selection is finalized.
The  $E_{\rm{extra}} < 0.5$ \gev region is defined as the nominal blinding
region which is slightly larger than the signal region for each mode.

For all the signal modes $E_{\rm{extra}}$ is optimized for the best signal significance
(assuming the branching fraction is $1\times 10^{-4}$).
The optimization yields to following requirements:

\begin{itemize}
\item For $\tautoe$: $\eextra <$0.31 \gev  
\item For $\tautomu$: $\eextra <$0.26 \gev 
\item For $\tautopi$: $\eextra <$0.48 \gev 
\item For $\tautorho$: $\eextra <$0.25 \gev 
\end{itemize}

\end{itemize}


\noindent The signal selection criteria for all signal modes are summarized in Table
\ref{tab:SigSelSummary}.

\begin{table}[!htb]
\caption{The selection criteria for different signal modes using a $\btodlnu$ tag are listed in this table.}
\vspace{-0.2cm}
\begin{center}
    \footnotesize
\begin{tabular}{|c|c|c|c|} \hline
$\tau^+ \to e^+ \nu_e \bar{\nu}_{\tau}$ & 
$\tau^+ \to \mu^+ \nu_\mu \bar{\nu}_{\tau}$ &
$\tau^+ \to \pi^+ \bar{\nu}_{\tau}$ &
$\tau^+ \to \pi^+ \pi^{0} \bar{\nu}_{\tau}$ \\ \hline \hline
$4.6 \le M_{miss} \le 6.7$ & $3.2 \le M_{miss} \le 6.1$ & $1.6 \le M_{miss}$ &   $M_{miss} \le 4.6$  \\ \hline
$p^{*}_{signal} \le 1.5$     &   --             & $1.6 \le p^{*}_{signal}$  & $1.7 \le p^{*}_{signal}$      \\ \hline
 \multicolumn{4}{|c|}{No IFR \KL\ } \\ \hline
$2.78 < R_{\tau\tau} < 4.0$ & $2.74 < R_{\tau\tau}$ & $2.84 < R_{\tau\tau}$ & $2.94 < R_{\tau\tau}$  \\ \hline
$m_{ee} > 0.1 \gevcc$ & & & \\ \hline
$N^{extra}_{\piz} \le 2$ & $N^{extra}_{\piz} \le 2$ & $N_{EMC \KL} \le 2$ & -- \\ \hline
     --        &   --         &  --      &  $\rho^{\pm}$ selection:                \\
                 &                &      &  0.64 $< M_{\rho^{\pm}}< $ 0.86 $\gev$   \\ 
                 &                &      &  $0.87 < \cos\theta_{\tau-\rho}$      \\ 
\hline 

$\eextra < 0.31$ $\gev$ & $\eextra < 0.26$ $\gev$ & $\eextra < 0.48$ $\gev$ &  $\eextra < 0.25$ $\gev$ \\ \hline
\end{tabular}
  \label{tab:SigSelSummary}
\end{center}
\end{table}
%


\subsubsection{Signal Efficiency}
\label{sec:SigEff}

The signal-side selection efficiencies 
for the $\tau$ decay modes are 
determined from signal MC simulation and summarized 
in Table~\ref{tab:signal_eff_dlnux}.
The signal efficiencies correspond to the number of events
selected in a specific signal decay mode, given that a tag $B$ has
been reconstructed.

\noindent \begin{table}[htb]
\caption{ The signal efficiencies, mode-by-mode, relative to the number of tags. The branching fraction for the 
given $\tau$ decay mode selected is included in the efficiency.
}
\vspace{0in}
\centering
\begin{tabular}{|l|c|} \hline 
Mode    &    Efficiency (BF Included) \\ \hline
$\tautoe$&	0.0414	$\pm$	0.0009 \\
$\tautomu$	&	0.0242	$\pm$	0.0007 \\
$\tautopi$    &	0.0492	$\pm$	0.0010 \\
$\tautorho$	&	0.0124	$\pm$	0.0005 \\
\hline
\end{tabular}
\label{tab:signal_eff_dlnux}
\end{table}

The selection efficiency for $\tautomununu$ is low compared to that of the
$\tautoenunu$ mode because the momentum spectrum 
of the signal muons peaks below 1 \gev/c, where the muon detection
efficiency is low. Since no minimum momentum requirement and no tight pion
identification criteria are applied to the
$\tautopinu$ signal selection, electron and muon signal tracks 
that fail particle identification requirement get selected in this mode. 
Any true $\tautopipiznu$ signal events, with a missed $\piz$ 
also get included in $\tautopinu$ selection mode. 
Therefore the $\tautopinu$ selection mode has the highest signal efficiency.

\subsection{Validation of Background Estimation from \boldmath{\eextra} Sidebands}
\label{sec:EextraSBExtrapolation}

We further study the agreement between simulation and data by using the
extra energy sideband region, and the ratio of the yields in this 
region to that in the signal region.
This is used mainly to test the reliability of 
the background estimation in the low $\eextra$ region 
by extrapolation from the higher $\eextra$ region.

The $\eextra > 0.5\gev$ region 
is defined as the ``sideband''~(sb). 
The ``signal region'' is defined separately for each selection mode.
For each control sample after applying appropriate selection cuts, 
the number of MC events in  the signal region ($N_{\mbox{\scriptsize{MC,Sig}}}$) 
and sideband ($N_{\mbox{\scriptsize{MC,sb}}}$) are counted
and their ratio ($R^{\mbox{\scriptsize{MC}}}$) is obtained. 

\begin{eqnarray}
R^{\mbox{\scriptsize{MC}}} & = & \frac{N_{\mbox{\scriptsize{MC,Sig}}}}{N_{\mbox{\scriptsize{MC,sb}}}}
\end{eqnarray}

\noindent Using the number of data events in the sideband ($N_{\mbox{\scriptsize{data,sb}}}$)
and the ratio $R^{\mbox{\scriptsize{MC}}}$, the number of expected background events in the 
signal region in data ($N_{\mbox{\scriptsize{exp,Sig}}}$) is estimated. 

\begin{eqnarray}
N_{\mbox{\scriptsize{exp,Sig}}} & = & N_{\mbox{\scriptsize{data,sb}}} \cdot R^{\mbox{\scriptsize{MC}}}
\end{eqnarray}

\noindent The number of expected data events ($N_{\mbox{\scriptsize{exp,Sig}}}$) 
in the signal region is compared with the observed number of data 
events ($N_{\mbox{\scriptsize{obs,Sig}}}$) in the signal region. The agreement between 
the above two quantities provide validation of background estimation in the 
low $\eextra$ region. 

Table~\ref{tab:EExtraSB_large} illustrates the level
of agreement between the sideband projections in MC and data. In general, the
agreement is at the $1\sigma$ level between the direct count in the MC
signal region and the projected data. The projections
in data are used to predict background for the final extraction, hence
we only rely on the data for this.

\noindent \begin{table}[htb]
\caption{%
The sideband-to-signalbox projection computed using a sideband region
where $\eextra > 0.5\gev$. The second column corresponds to the ratio
of yields in the signal region and sideband as measured in MC.
}
\vspace{0in}
\footnotesize
\centering
\begin{tabular}{|c|c|c|c|c|} \hline
Mode      & ratio (MC)        &  upper sb (Data)   &  signal region (Proj)&signal region (MC)  \\
\hline
\hline
electron  & 0.137  $\pm$ 0.015  &  305.00 $\pm$ 17.46  &  41.91  $\pm$ 5.19     &39.72  $\pm$ 4.07 \\
\hline
muon      & 0.037  $\pm$ 0.004  &  965.00 $\pm$ 31.06  &  35.39  $\pm$ 4.16     &36.13  $\pm$ 4.02\\
\hline
pion      & 0.043  $\pm$ 0.004  &  2288.00 $\pm$ 47.83 &  99.09  $\pm$ 9.10     &87.69  $\pm$ 7.72\\
\hline
rho       & 0.005  $\pm$ 0.001  &  2805.00 $\pm$ 52.96 &  15.30  $\pm$ 3.48     &15.81  $\pm$ 3.58\\
\hline
\end{tabular}
\label{tab:EExtraSB_large}
\end{table}

\section{VALIDATION OF TAG \boldmath{B} YIELD AND \boldmath{\eextra} SIMULATION}
\label{sec:EextraValidation}

The tag $B$ yield and $E_{\rm{extra}}$ distribution in signal and background MC simulation 
are validated using various control samples.
The level of agreement between the data and simulation distributions provides
validation of the $E_{\rm{extra}}$ modeling in the simulation and corrects
for differences in the yield of reconstructed tag $B$'s.

``Double-tagged'' events, for which
both of the $B$ mesons are reconstructed in tagging modes, 
$\btodlnux$~vs. $\B^+ \to \bar{D}^{0} \ell^+ \nul X$ are used as the main control sample. 
Due to the large branching fraction and high tagging efficiency for these events, 
a sizable sample of such events is reconstructed in the on-resonance dataset. 
Due to all of the decay products of the $\Y4S$ being correctly accounted for the double-tagged events 
reconstructed have a high purity. 

To select double-tag events 
we require that the two tag $B$ candidates do not share 
any tracks or neutrals. If there are more than two such 
non-overlapping tag $B$ candidates in the event then the 
best candidates are selected as those with the largest $\Dz$-$\ell$ vertex probability, as with the signal search.
The number of double-tagged events ($N_{2}$) is given by
\begin{eqnarray}
\label{eqn:dbltageff}%
N_2 & = & \varepsilon^{2} N.
\end{eqnarray}
where $N$ is the number of $B\bar{B}$ events in the sample and $\varepsilon$ is the tag efficiency that is compared
between data and MC.
Using the expression in equation~\ref{eqn:dbltageff} we calculate the efficiencies $\varepsilon_{\rm{data}}$
and $\varepsilon_{\rm{MC}}$. 
The correction factor, ratio of the efficiencies between data and simulation, from this method is given in equations~\ref{correction_runs1to3},~\ref{correction_run4}~and~\ref{correction_run5} for Runs~1--3, Run~4 and Run~5 respectively.
\begin{eqnarray}
\frac{\varepsilon_{\mbox{\scriptsize{Runs~1--3}}}}{\varepsilon_{\mbox{\scriptsize{MC}}}} & = 1.05 \pm 0.02
\label{correction_runs1to3}
\end{eqnarray}
\begin{eqnarray}
\frac{\varepsilon_{\mbox{\scriptsize{Run~4}}}}{\varepsilon_{\mbox{\scriptsize{MC}}}} & =  1.00 \pm 0.03
\label{correction_run4}
\end{eqnarray}
\begin{eqnarray}
\frac{\varepsilon_{\mbox{\scriptsize{Run~5}}}}{\varepsilon_{\mbox{\scriptsize{MC}}}} & =  0.97 \pm 0.03
\label{correction_run5}
\end{eqnarray}
\noindent It was directly verified that data taken during Runs~1--3 agreed in both shape 
and normalized yield whereas during Run~4 and Run~5 data were taken with the machine operating 
in a mode of continuous injection which may affect detector backgrounds differently. These 
runs are therefore considered separately.

\begin{figure}[htb]
     \centering
     \subfigure{
          \label{fig:Eextra-DoubleTag-run1-3}
          {\includegraphics[width=.3\textwidth]{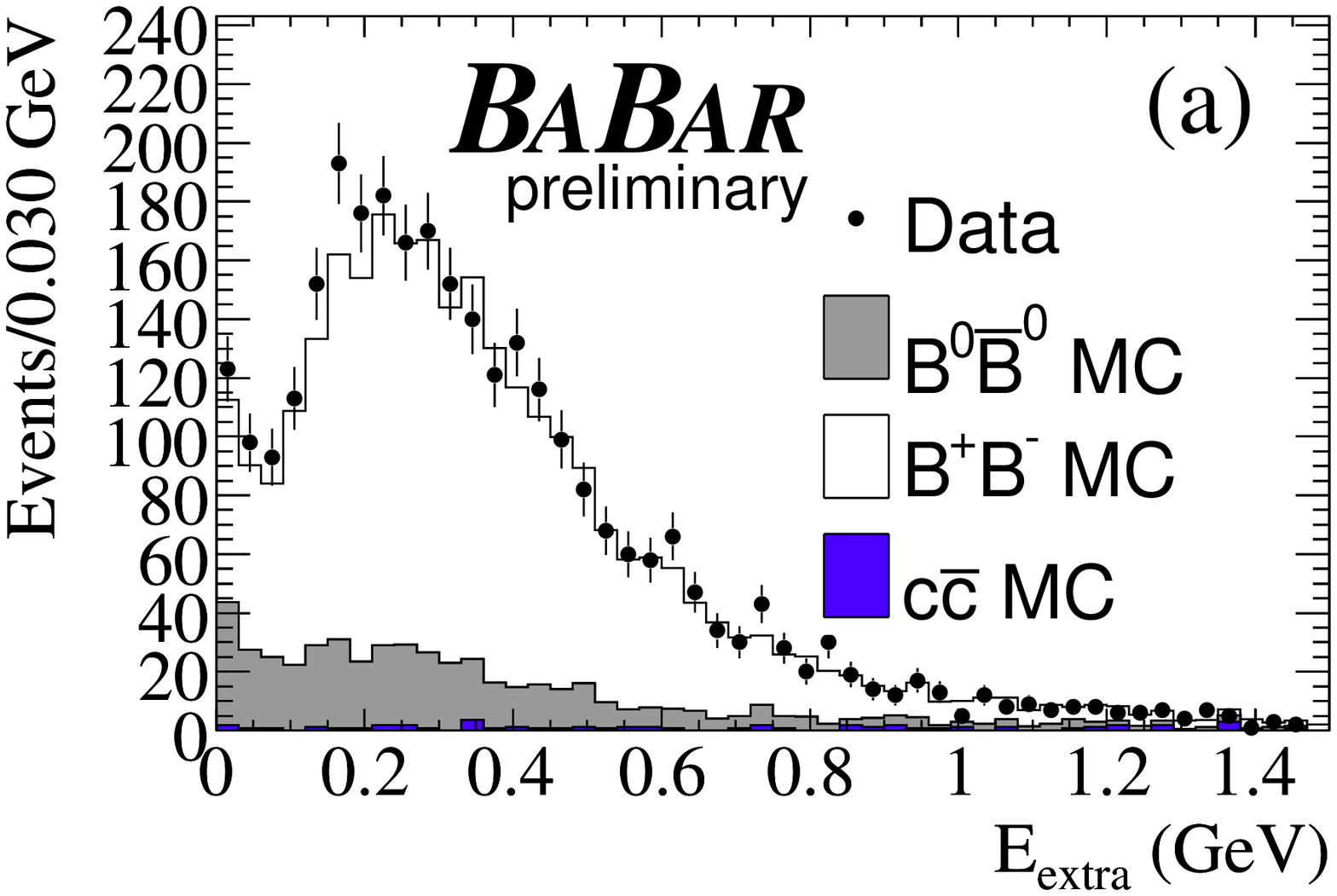}}}
     \subfigure{
          \label{fig:Eextra-DoubleTag-run4}
          {\includegraphics[width=.3\textwidth]{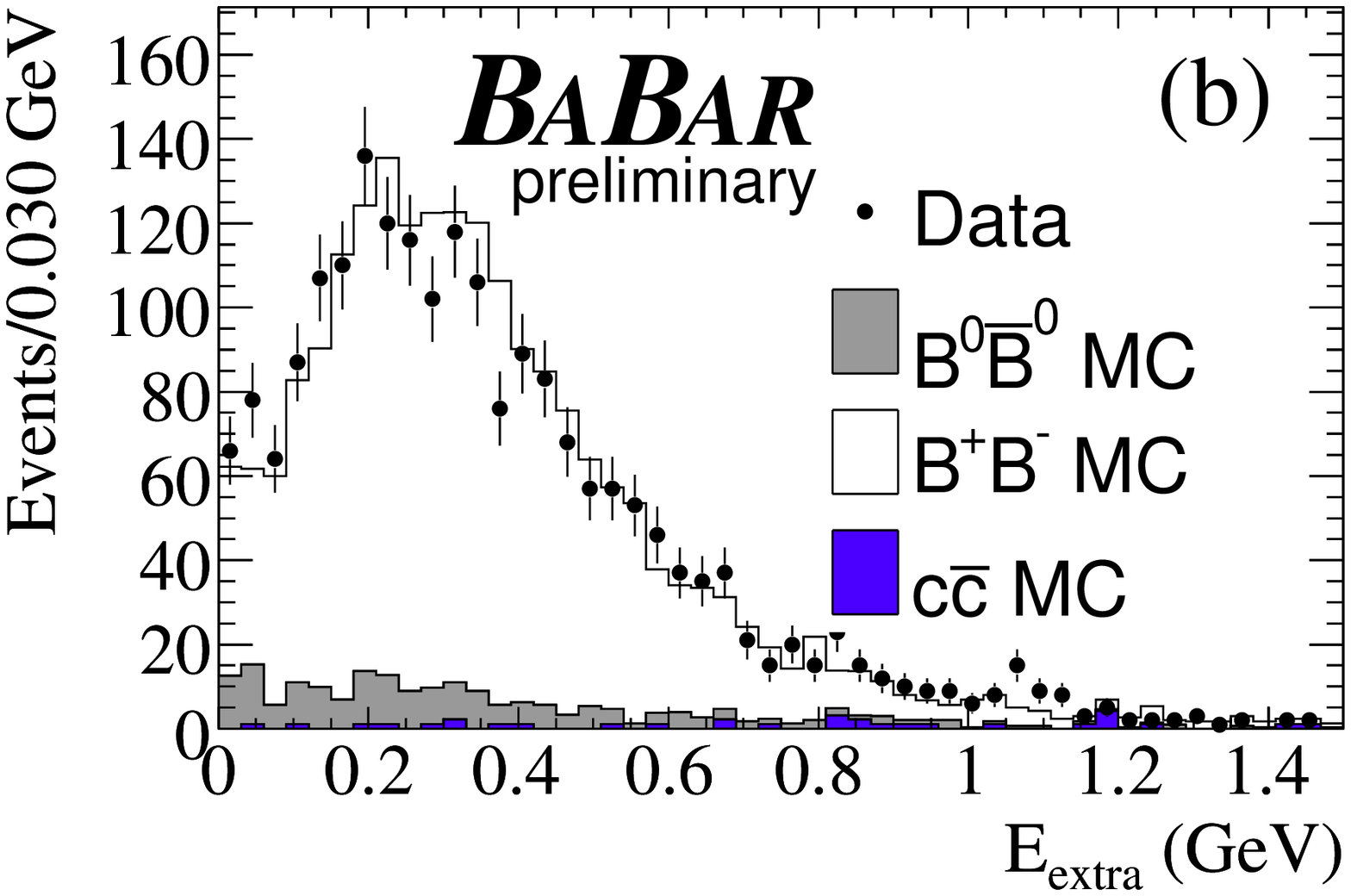}}} 
     \subfigure{
          \label{fig:Eextra-DoubleTag-run5}
          {\includegraphics[width=.3\textwidth]{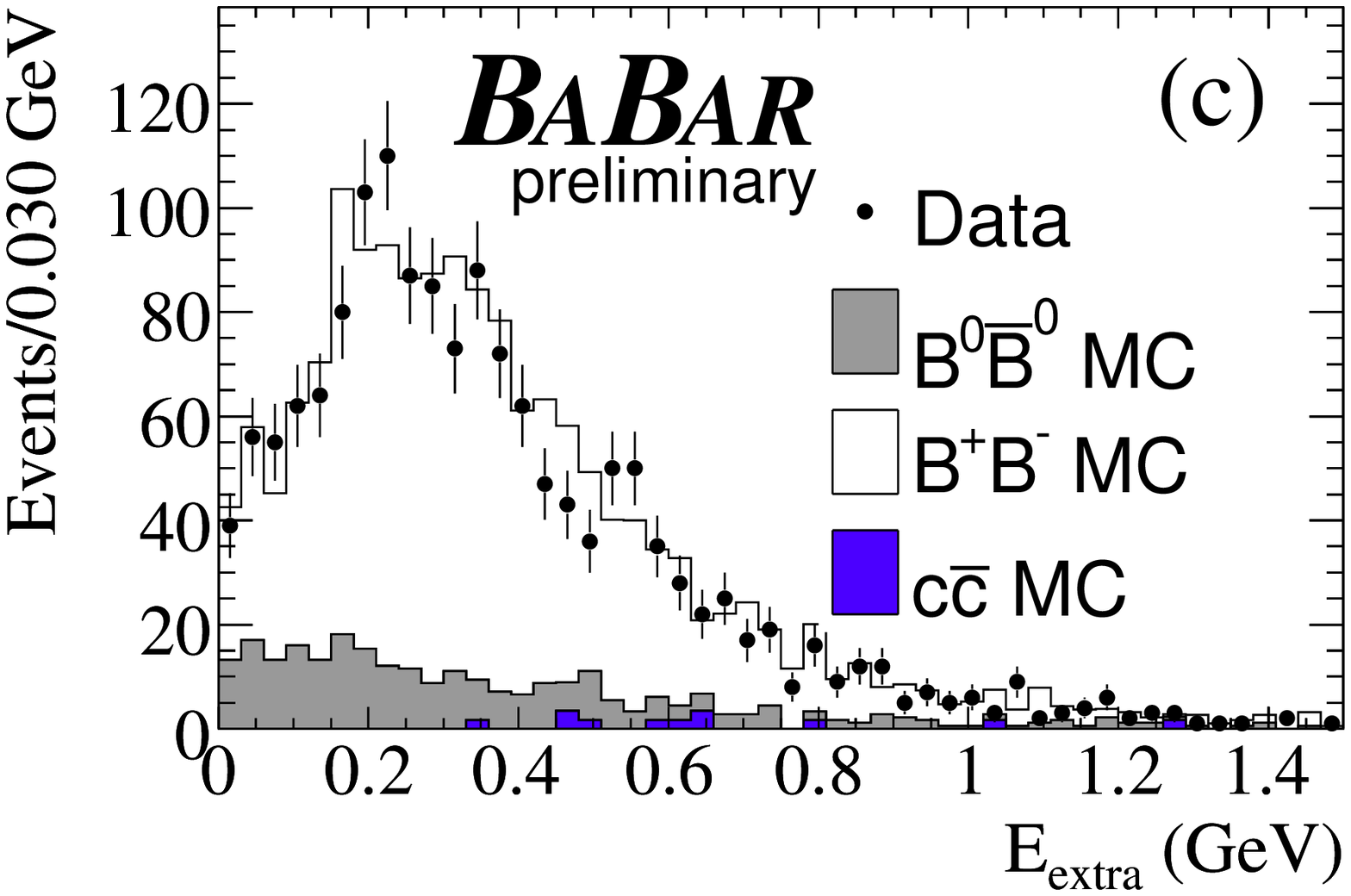}}} 
\vspace{-0.4cm}
     \caption{The distribution of the remaining neutral energy ($\eextra$) for double-tagged events, plotted for 
generic MC and data: a) Runs 1-3, b) Run 4 and c) Run 5. No off-resonance data events
are seen in the $\eextra$ region plotted here. In these events both of the $D^{0} \ell$ candidates from 
double-tag are required to pass the selection described in section~\ref{sec:TagReco} 
and best candidate selection. 
The differences in these distributions are used for obtaining the systematic error for 
tagging efficiency correction.}
     \label{fig:DoubleTagEextra}
\end{figure}

The $E_{\rm{extra}}$ for the double-tagged sample is calculated by summing the 
CM energy of the photons which are not associated with 
either of the tag $B$ candidates. The sources of neutrals contributing
to the $E_{\rm{extra}}$ distribution in double-tagged events  
are similar to those contributing to the $E_{\rm{extra}}$ distribution  
in the signal MC simulation. Therefore the agreement of
the $E_{\rm{extra}}$ distribution between data and MC simulation for the 
double-tagged sample, in figure \ref{fig:DoubleTagEextra},
is used as a validation of the $E_{\rm{extra}}$ simulation in the signal 
MC.  

The simulation is further validated by comparing a sample of events where the signal candidate and
tag $B$ candidate are of the ``wrong-sign'' with non-zero net charge. The agreement between
data and simulation for all signal modes for the background estimation in the
$E_{\rm{extra}}$ signal region provides a useful cross-check.

\section{STUDIES OF SYSTEMATICS}
\label{sec:Systematics}

The main sources of uncertainty in the determination of the $\btn$
branching fraction are the following:

\begin{itemize}
\item Uncertainty in tagging efficiency determination 

\item Uncertainty in determination of the efficiency $\varepsilon_{i}$
for each selection mode.

\item Uncertainty in the determination of the number of expected background
events in the signal region for each selection mode.

\end{itemize}

A small uncertainty of 1.1\% also enters the branching ratio limit calculation 
from the estimation of the number of $B^{+}B^{-}$ events present in the 
data sample \cite{ref:BCount}.
The systematic uncertainties are summarized in table~\ref{tab:SignalEffSys}.

\subsection{Tagging Efficiency Systematics}
\label{sec:tagEffSys}

The tagging efficiency and yield in signal simulation is corrected 
using the double-tagged events. The selection of 
double-tagged events is described
in section~\ref{sec:EextraValidation}.

We take the 1.9\%, 3.0\% and 3.1\% 
errors (from equations~\ref{correction_runs1to3},~\ref{correction_run4}~and~\ref{correction_run5}) 
obtained from the double tag method as the systematic uncertainties
associated with the tagging efficiency and yield correction in MC.
The combined, luminosity weighted, tag $B$ yield systematic uncertainty  
is 1.5\%.
The luminosity weighted tag $B$ yield correction is 1.01.

\subsection{\boldmath{\eextra} Systematic Uncertainty}
\label{sec:EExtraSys}

The systematic uncertainty due to the mis-modeling of the $\eextra$ variable is extracted 
using the double-tagged events. The selection of 
double-tagged events is described
in Section~\ref{sec:EextraValidation}. 
A cut is imposed on the $\eextra$ distributions shown in Figures~\ref{fig:Eextra-DoubleTag-run1-3},
\ref{fig:Eextra-DoubleTag-run4}~and~\ref{fig:Eextra-DoubleTag-run5} to extract the yield of candidates
satisfying $\eextra <$~0.5GeV. This yield is then compared to the number of candidates in the full sample.
Comparing the ratio extracted from MC to that extracted from data yields a correction factor, the error
on which is taken as the systematic uncertainty for $\eextra$. 
These values are broken up by run and we extract the following numbers: 
Runs~1--3 = 0.98$\pm$0.06, Run~4 = 0.99$\pm$0.06, Run~5 = 1.02$\pm$0.08
The combined, luminosity weighted systematic uncertainty for $\eextra$ is 3.8\%.
The luminosity weighted $\eextra$ correction is 0.99.
 
\subsection{Uncertainties in the signal selection efficiencies in each selection mode}
\label{sec:sigEffSys}

Besides the tagging efficiency uncertainty,
the contribution to the systematic uncertainties in the determination of the 
efficiencies comes from systematic uncertainty on the tracking efficiency,
particle identification, and simulation of the neutral clusters in the 
calorimeter which contribute to the $E_{\rm{extra}}$ distribution, and $\KL$ identification. 
The different contributions to the systematic uncertainty on the selection efficiencies are
listed in table \ref{tab:SignalEffSys}.

\begin{table}[!htb]
\caption{Contribution to the systematic uncertainty on the signal selection efficiencies in different selection modes.
These uncertainties are added together in quadrature with the uncertainty on the tag $B$ yield, extracted from the
double-tagged control sample, of 1.5\%. The uncertainty on MC statistic is added in quadrature to obtain the total
systematic uncertainty.}
\begin{center}
\begin{tabular}{|c|c|c|c|c|c|c|c|} \hline
Selection     &  tracking & Particle       & $\KL$  &   $\eextra$  & $\piz$        & Total       &  Correction \\
modes         &   (\%)    & Identification &        &   modeling   & modeling      & Systematic  &  Factor      \\
              &           &      (\%)      & (\%)   &     (\%)     &  (\%)         & Error (\%)  &              \\
\hline \hline
$\enunu$      &  0.3     &  2.0            &  3.6    &   3.8      &  --       &     5.8       &  0.982   \\ \hline
$\mununu$     &  0.3     &  3.0            &  3.6    &   3.8      &  --       &     6.2       &  0.893    \\ \hline
$\pinu$       &  0.3     &  1.0            &  6.2    &   3.8      &  --       &     7.5       &  0.966    \\ \hline
$\pipiznu$    &  0.3     &  1.0            &  3.6    &   3.8      &  1.8      &     5.8       &  0.961    \\ \hline
\end{tabular}
\end{center}
\label{tab:SignalEffSys}
\end{table}

\subsection{Uncertainties on $\KL$ modeling}
\label{sec:KLSys}

The systematic uncertainty on the modeling of $\KL$ candidates is extracted using the double-tagged events
outlined in section~\ref{sec:EextraValidation}.
A comparison between data and simulation is used to extract 
both a correction and a systematic uncertainty, similarly to the method used for $\eextra$. 
We quantify this comparison by comparing the yield with a cut demanding exactly zero reconstructed 
IFR measured $\KL$ candidates remaining, with a sample where any number of $\KL$ candidates remain and 
take the ratio of ratios from the MC and data.
We extract the following values for corrections and systematic uncertainties: 
Runs~1--3 = 0.98$\pm$0.05, Run~4~=~1.00$\pm$0.06, Run~5~=~0.98$\pm$0.08, hence percentage uncertainties of
5.1\%, 6.0\% and 8.2\%. 
The correction factors are all close to unity as expected. 
The combined, luminosity weighted systematic uncertainty for IFR $\KL$ candidates is 3.6\%.
The luminosity weighted IFR $\KL$ correction is 0.99.

The same exercise is performed for $\KL$ candidates reconstructed in the EMC. 
We extract the following values for corrections and systematic uncertainties: 
Runs~1--3~=~0.88$\pm$0.05, Run~4~=~1.00$\pm$0.10, Run~5~=~1.08$\pm$0.11.
Percentage uncertainties are 5.7\%, 10\% and 10.2\%.
The combined, luminosity weighted systematic uncertainty for EMC $\KL$ candidates is 5.1\%.
The luminosity weighted EMC $\KL$ correction is 0.97.

\section{RESULTS}
\label{sec:Physics}

\begin{figure}[htb]
\begin{center}
\hspace{-0.1in}
     \subfigure{\includegraphics[width=.46\textwidth]{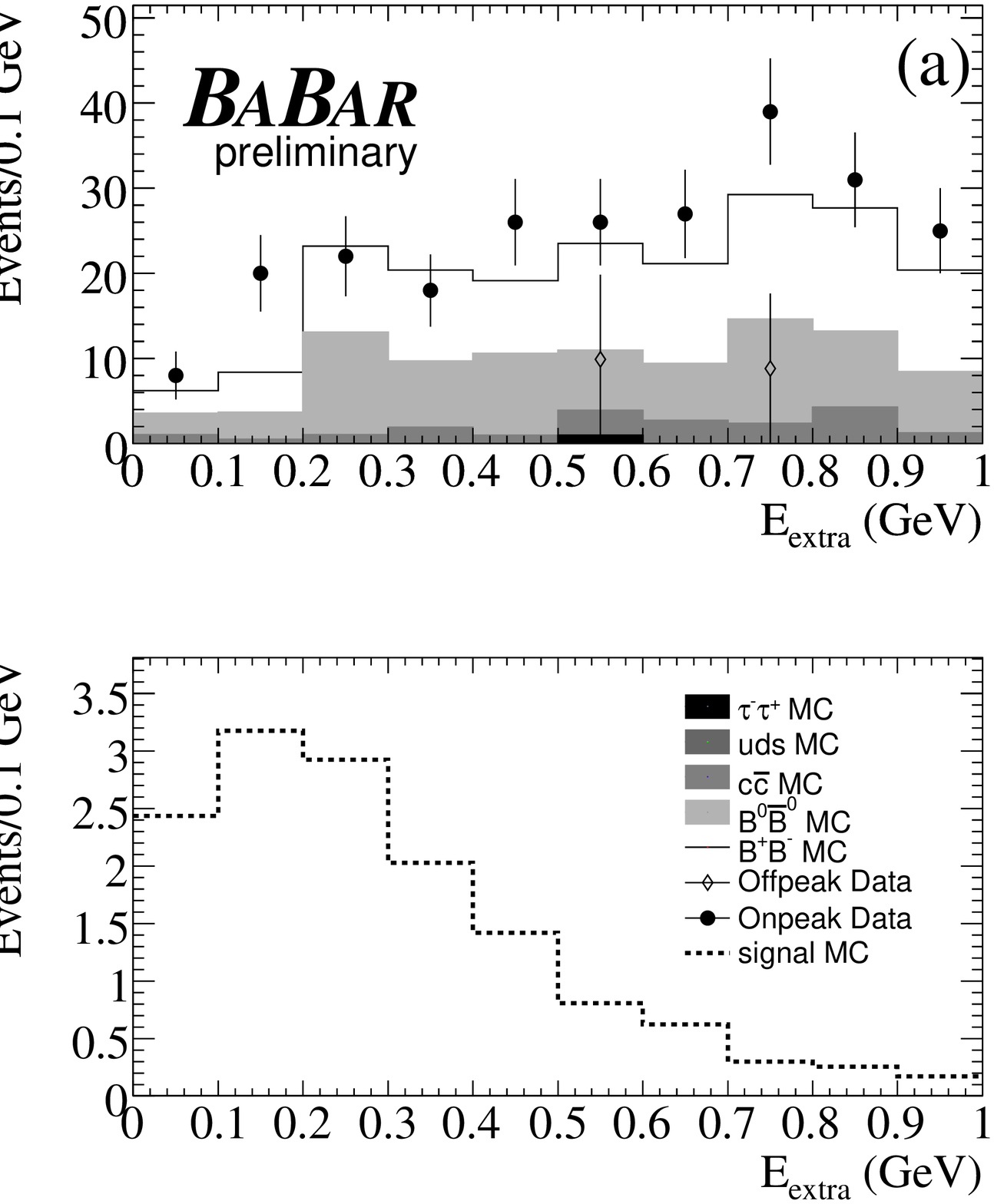}}
     \hspace{.02in}
      \subfigure{\includegraphics[width=.46\textwidth]{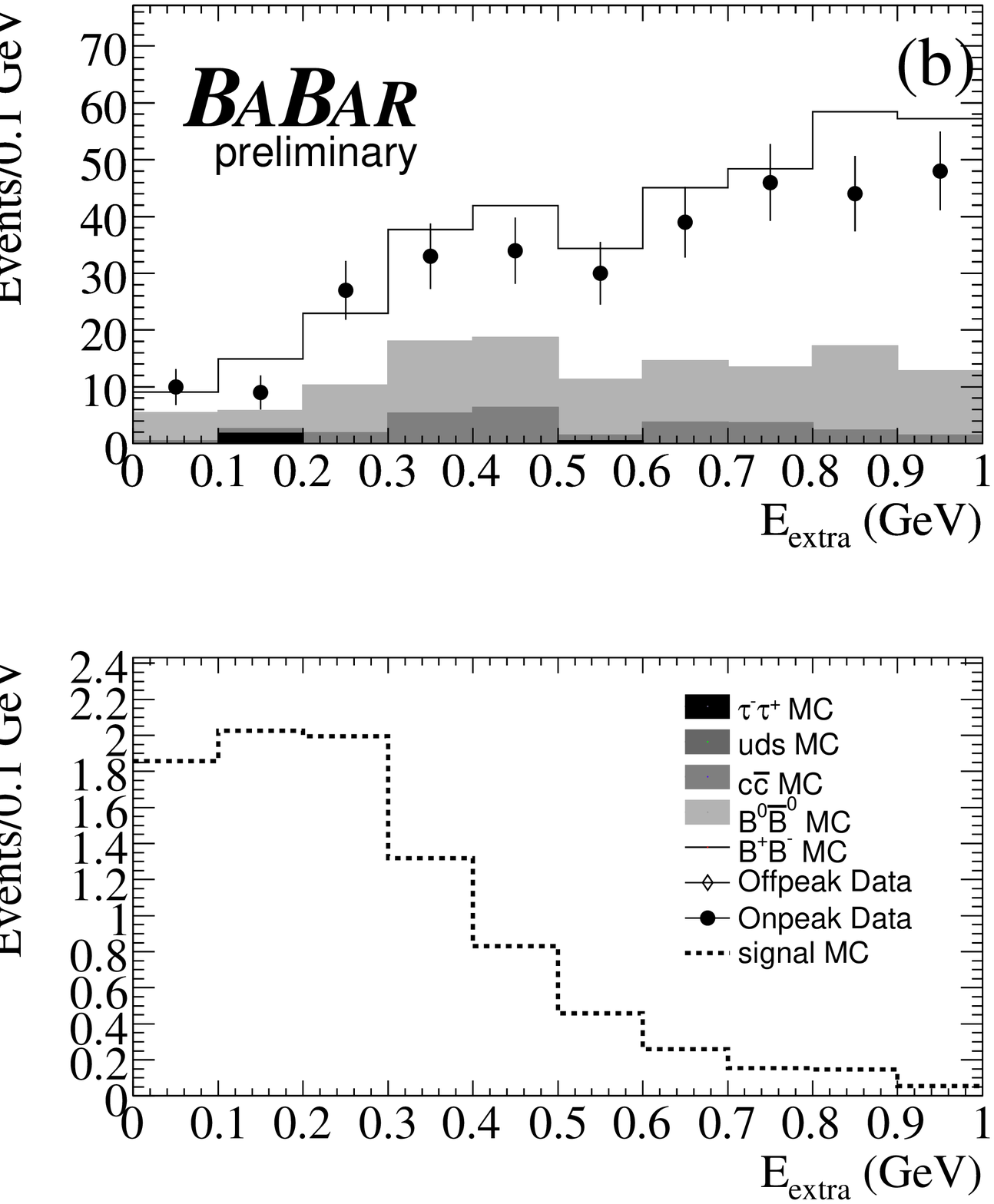}}
\end{center}
\vspace{-1cm}
\caption{Total extra energy is plotted after all cuts have been applied in 
          the mode (a)~\tautoe~and (b) \tautomu.
	  Off-resonance data and MC have been normalized to the on-resonance luminosity. 
	  Simulated $\btn$ signal MC is plotted (lower) for comparison. 
      }
\label{fig:eextra_allcuts_blind_e}
\end{figure}

\begin{figure}[htb]
\begin{center}
     \subfigure{\includegraphics[width=.46\textwidth]{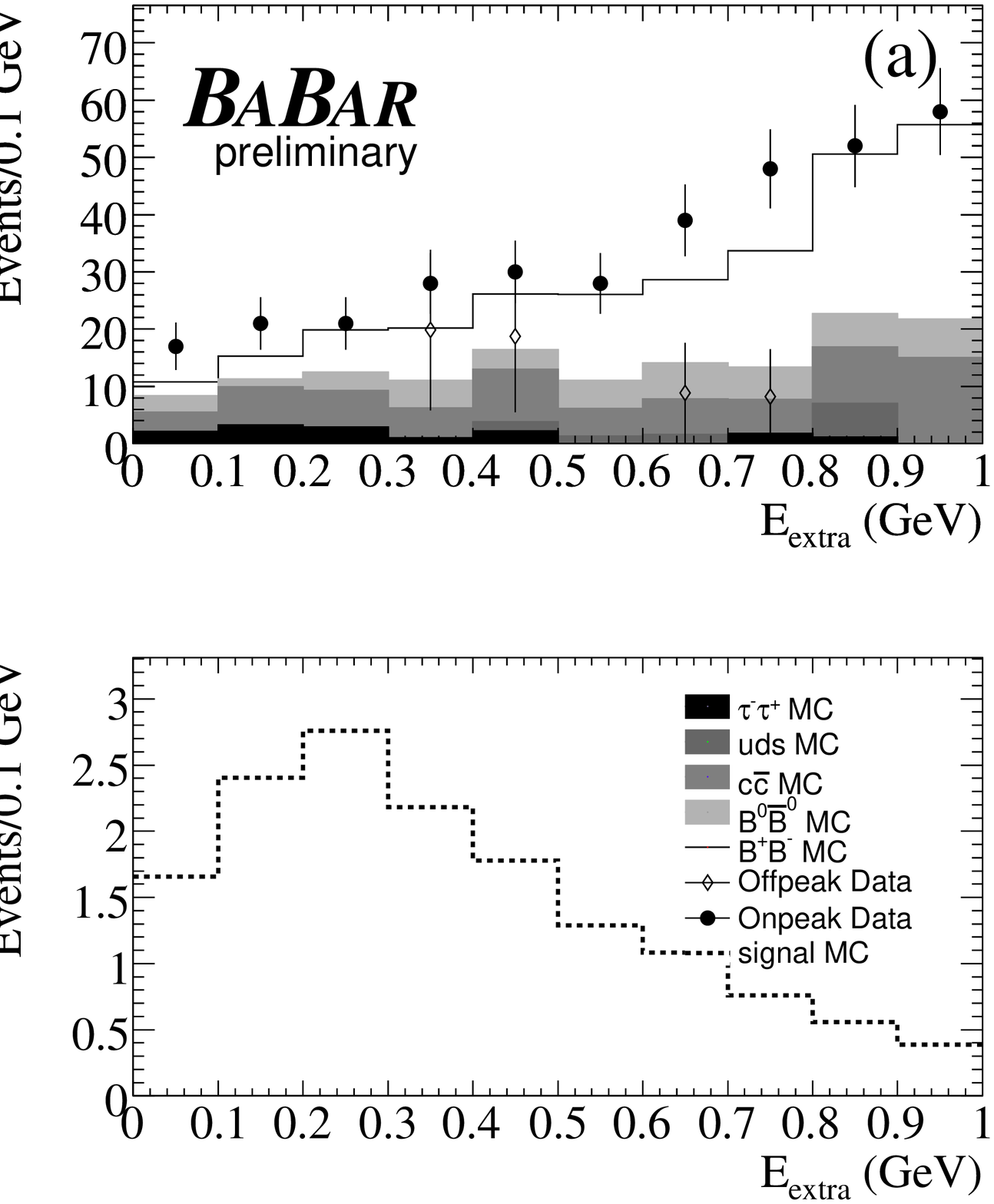}}
     \hspace{.02in}
      \subfigure{\includegraphics[width=.46\textwidth]{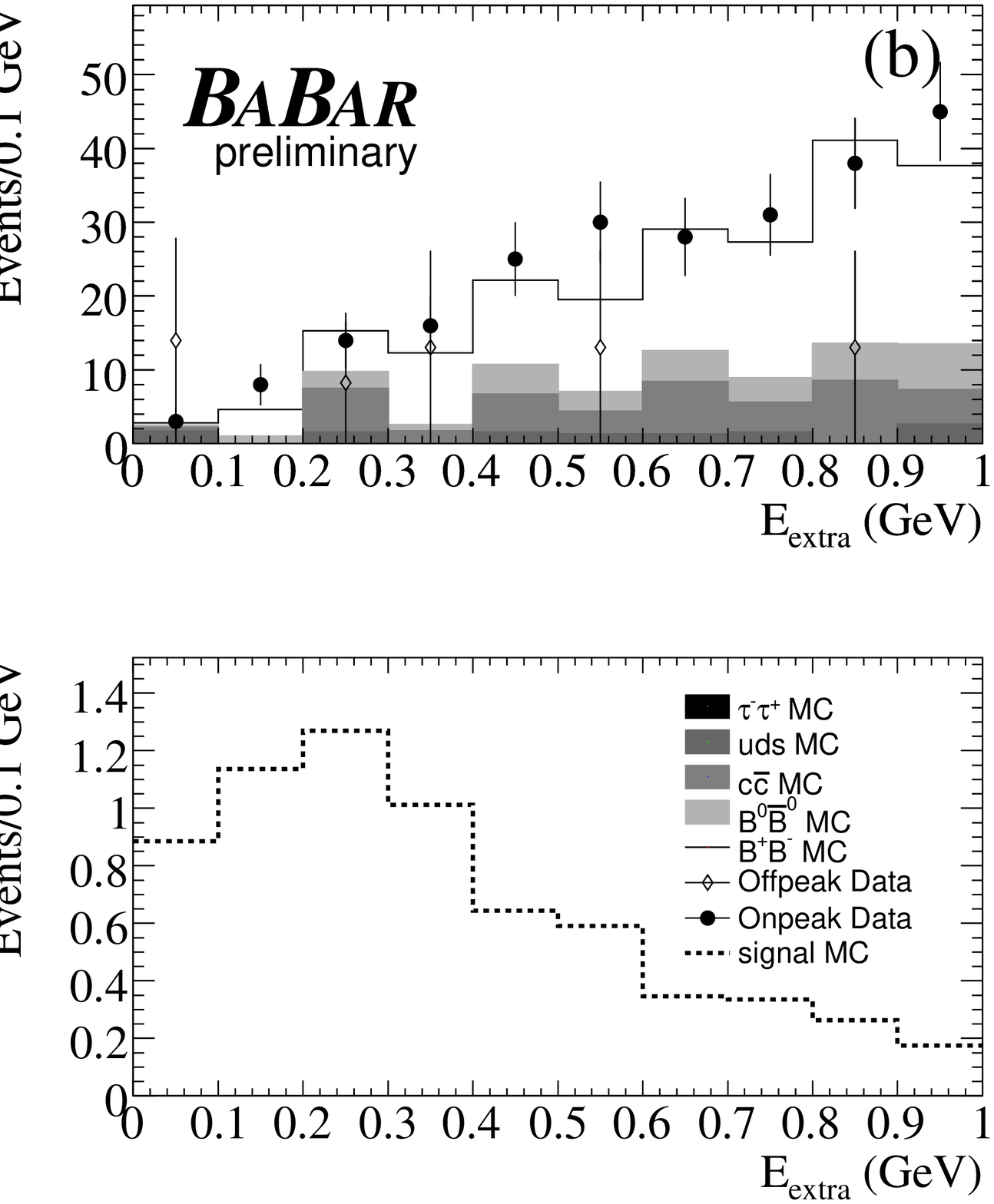}}
\end{center}
\vspace{-1cm}
\caption{Total extra energy is plotted after all cuts have been applied in 
          the mode (a) \tautopi~and~(b) \tautorho.
	  Off-resonance data and MC have been normalized to the on-resonance luminosity. 
	  Simulated $\btn$ signal MC is plotted (lower) for comparison. 
      }
\label{fig:eextra_allcuts_blind_mu}
\end{figure}

\begin{figure}[htb]
\begin{center}
\includegraphics[width=.5\textwidth]{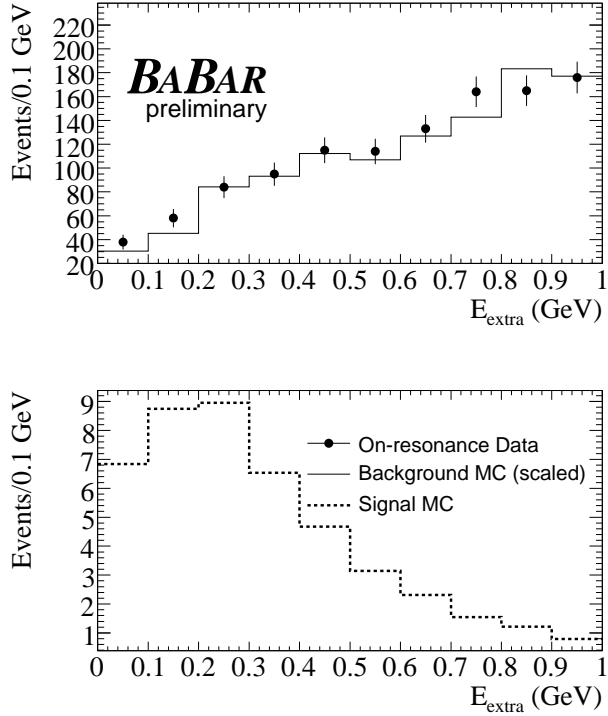}
\end{center}
\vspace{-1cm}
\caption{Total extra energy is plotted after all cuts have been applied with all modes combined. Off-resonance data and MC have been normalized to the on-resonance luminosity. Events in this distribution
	  are required to pass all selection criteria. In addition the background MC have been scaled according
to the ratio of predicted backgrounds from data and MC as presented in section~\ref{sec:EextraSBExtrapolation}.
	  Simulated $\btn$ signal MC is plotted (lower) for comparison. 
      }
\label{fig:eextra_allcuts_ALL}
\end{figure}

\begin{figure}[htb]
\begin{center}
\hspace{-0.1in}
\subfigure{\includegraphics[%
				     width=0.48\linewidth,
				     height=3in,
				     keepaspectratio]{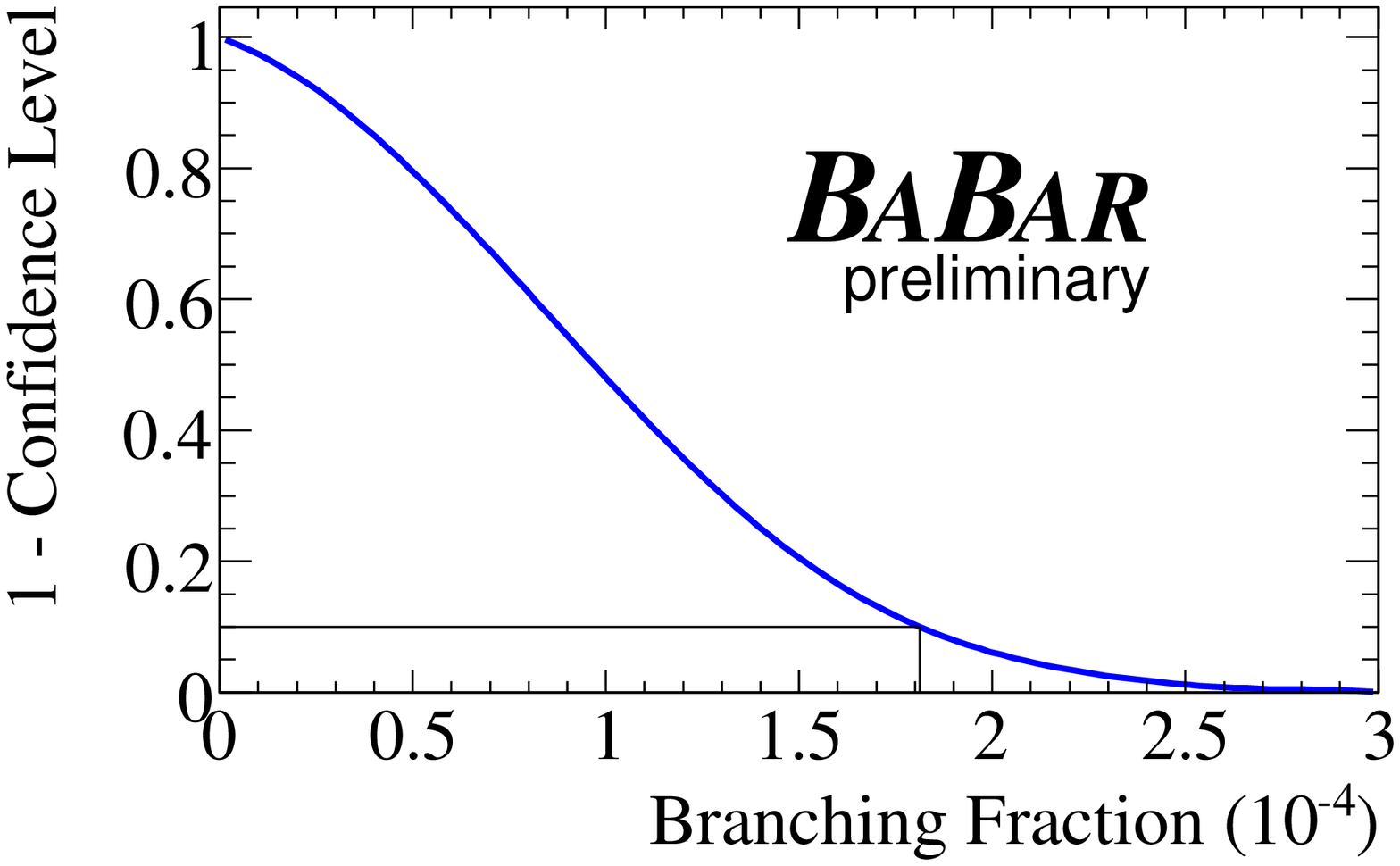}}
\subfigure{\includegraphics[%
				     width=0.48\linewidth,
				     height=3in,
				     keepaspectratio]{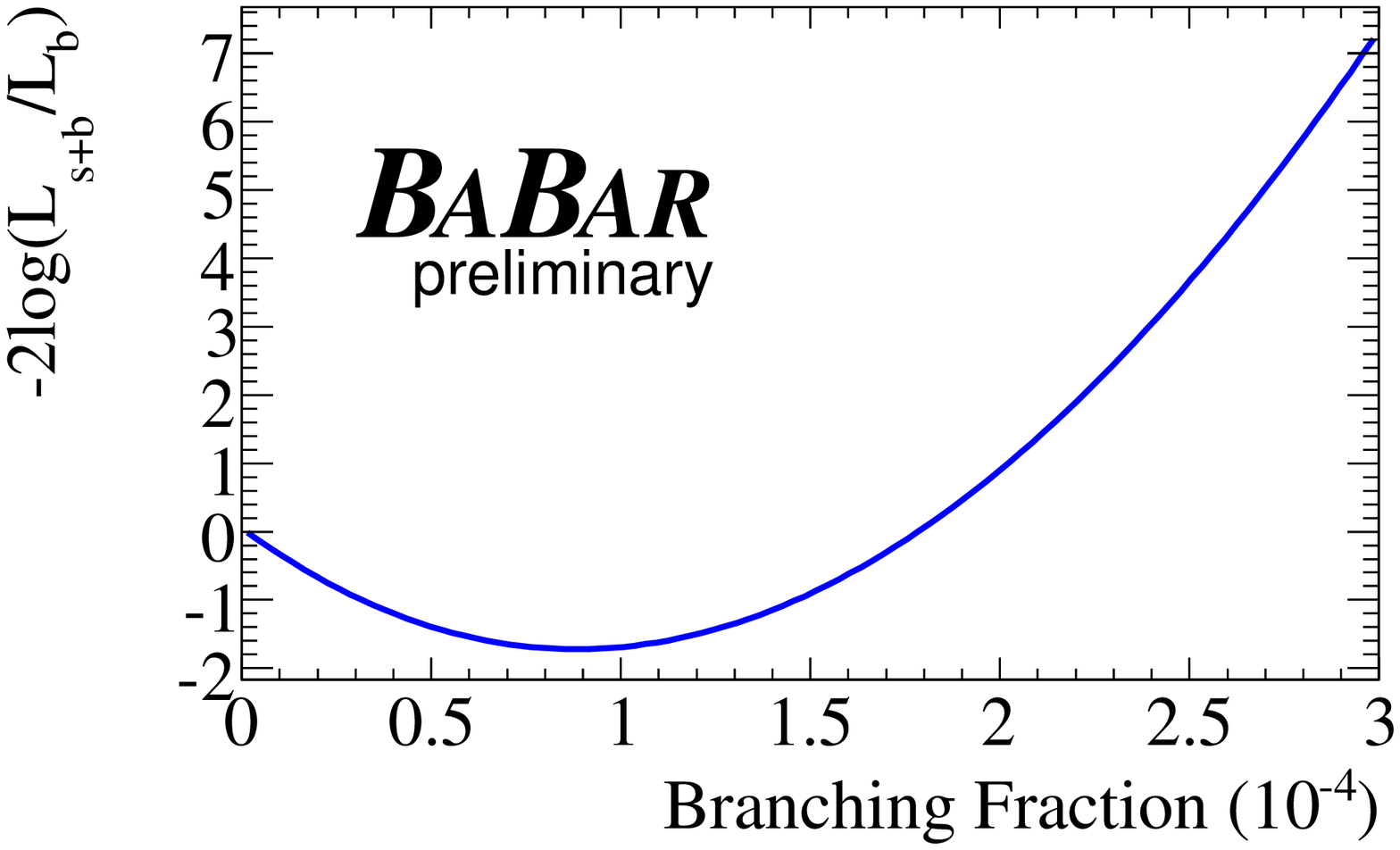}}
\end{center}
\vspace{-1cm}
\caption{The confidence level vs branching fraction is shown (left) to illustrate the extracted upper limit.
The negative log likelihood curve (right) illustrates the central value and it's corresponding uncertainty.
      }
\label{fig:unblindCL}
\end{figure}

After finalizing the signal selection criteria, the signal region
in the on-resonance data is examined. 
Table~\ref{tab:unblind-result} lists the
number of observed events in on-resonance data in the signal region,
together with the expected number of background events in the 
signal region.
Figures~\ref{fig:eextra_allcuts_blind_e}~and~\ref{fig:eextra_allcuts_blind_mu} show the $E_{\rm{extra}}$ 
distribution in data and simulation for each of the $\tau$ decay modes considered.
Data is overlayed on the summed MC contribution, scaled to the dataset luminosity, and signal
MC is plotted for comparison. 
Figure~\ref{fig:eextra_allcuts_ALL} shows the $E_{\rm{extra}}$ distribution for all modes combined. 
\begin{table}[hbt]
\centering
\caption{\label{tab:unblind-result}The observed number of on-resonance data events in the signal region are shown, together with number of expected background events. The background estimations include systematic corrections referred to in section~\ref{sec:EextraSBExtrapolation}.}
\begin{tabular}{|c|c|c|} \hline
Selection    & Expected           &  Observed Events  \\ 
             & Background Events  &  in On-resonance Data  \\ \hline
$\enunu$     & 41.9 $\pm$ 5.2  & 51  \\ \hline
$\mununu$    & 35.4 $\pm$ 4.2  & 36  \\ \hline
$\pinu$      & 99.1 $\pm$ 9.1  & 109  \\ \hline
$\pipiznu$   & 15.3 $\pm$ 3.5  & 17  \\ \hline
\hline
All modes    & 191.7 $\pm$ 11.8 & 213  \\ \hline
\end{tabular}
\end{table}
We determine the \btn \xspace branching fraction from the number of signal
candidates $s_i$ in data for each $\tau$ decay mode, according to $s_i =
N_{B \overline{B}}  \mathcal{B}(\btn) \varepsilon_{\rm{tag}} \varepsilon_i$. Here 
$N_{B \overline{B}}$ is the total number of $B\overline{B}$ pairs in data,
$\varepsilon_{\rm{tag}}$ is the tag reconstruction efficiency in signal 
MC; $\varepsilon_i$ is the signal-side selection efficiency in
different $\tau$ decay modes calculated with respect to the total number
of reconstructed tag $B$ mesons. Table \ref{tab:EffBcountSummary} shows 
the values of $N_{B \overline{B}}$, $\varepsilon_{\rm{tag}}$ and 
$\varepsilon_i$ after applying appropriate systematic corrections 
(see section~\ref{sec:Systematics}).
\begin{table}[htb]
\centering
\caption{The corrected tag and signal efficiencies. Two errors are quoted:
the first is the MC statistical uncertainty, and the second is the 
systematic error computed from the sources in section~\ref{sec:Systematics}.}
\label{tab:EffBcountSummary}
\vspace{0.1cm}
\begin{tabular}{|c|c|c|}
\hline
Efficiency        &     Corrected         &      Relative Systematic Error (\%) \\
\hline\hline
Tag               &     $(6.77 \pm 0.05(\mbox{stat.}) \pm 0.10(\mbox{syst.}))\times10^{-3}$ & 1.5 \\
\hline
$\varepsilon(\tautoe)$ & $(4.06 \pm 0.09(\mbox{stat.}) \pm 0.23(\mbox{syst.}))\times10^{-2}$ &  5.6\\
\hline
$\varepsilon(\tautomu)$ & $(2.16 \pm 0.06(\mbox{stat.}) \pm 0.13(\mbox{syst.}))\times10^{-2}$ & 6.0 \\
\hline
$\varepsilon(\tautopi)$ & $(4.88 \pm 0.10(\mbox{stat.}) \pm 0.35(\mbox{syst.}))\times10^{-2}$ & 7.3 \\
\hline
$\varepsilon(\tautopipiz)$ & $(1.16 \pm 0.05(\mbox{stat.}) \pm 0.07(\mbox{syst.}))\times10^{-2}$ & 5.6 \\\hline
\end{tabular}
\end{table}
The results from each decay mode are combined using the ratio 
$Q = {\cal L}(s+b)/{\cal L}(b)$,
where ${\cal L}(s+b)$ and ${\cal L}(b)$ are the
likelihood functions for signal plus background and background-only
hypotheses, respectively~\cite{ref:cls}:
\begin{equation}
  {\cal L}(s+b) \equiv
  \prod_{i=1}^{n_{ch}}\frac{e^{-(s_i+b_i)}(s_i+b_i)^{n_i}}{n_i!},
        \;
  {\cal L}(b)   \equiv
  \prod_{i=1}^{n_{ch}}\frac{e^{-b_i}b_i^{n_i}}{n_i!},
  \label{eq:lb}
\end{equation}
We include the statistical and systematic uncertainties on the expected
background ($b_{i}$) in the likelihood definition by
convolving it with a Gaussian distribution ($\mathcal{G}$).
The mean of $\mathcal{G}$ is $b_{i}$, and
the standard deviation ($\sigma_{b_{i}}$) of $\mathcal{G}$ is the  
statistical and systematic errors on $b_i$ added 
in quadrature~\cite{bib:lista},
\begin{equation}
\mathcal{L}(s_i+b_i) \rightarrow \mathcal{L}(s_i+b_i) \otimes \mathcal{G}(b_i,\sigma_{b_{i}})%
\;
\end{equation}
(similarly for $\mathcal{L}(b_i)$).
\noindent The results from this procedure are illustrated in Figure~\ref{fig:unblindCL}.

\noindent We determine the following branching fraction
\begin{equation}
\mathcal{B}(\btn) = (0.88^{+0.68}_{-0.67}(\mbox{stat.}) \pm 0.11 (\mbox{syst.})) \times 10^{-4},
\label{eqn:bf}
\end{equation}
\noindent and also set an upper limit at the 90\% confidence level of
\begin{equation}
\mathcal{B}(\btn) < 1.8 \times 10^{-4}. 
\label{eqn:ul}
\end{equation}
\noindent Figure~\ref{fig:unblindCL} shows the distributions of confidence level vs branching fraction 
 and the negative log likelihood curve illustrating the extracted upper limit and central 
value respectively.

\noindent Using the measured central value for $\mathcal{B}(\btn)$ and taking the known values of
$G_F$, $m_B$, $m_{\tau}$ and $\tau_{B}$ from Ref.~\cite{ref:pdg2004} we calculate, from equation~\ref{eqn:br}, 
the product of the $B$ meson decay constant and $|\Vub|$ to be
$f_{B}\cdot|\Vub| = (7.0^{+2.3}_{-3.6}(\mbox{stat.})^{+0.4}_{-0.5}(\mbox{syst.}))\times10^{-4}$~GeV.

\section{SUMMARY}
\label{sec:Summary}
We have performed a search for the decay
process \btn. To accomplish this a sample of
semileptonic $B$ decays (\dlnux) has been used to
reconstruct one of the $B$ mesons and the remaining information in
the event is searched for evidence of \btn. 
A branching fraction of
\begin{equation}
\mathcal{B}(\btn) = (0.88^{+0.68}_{-0.67}(\mbox{stat.}) \pm 0.11 (\mbox{syst.})) \times 10^{-4},
\label{eqn:finalbf}
\end{equation}
\noindent is measured and we set an upper limit at the 90\% confidence level of
\begin{equation}
\mathcal{B}(\btn) < 1.8 \times 10^{-4}. 
\label{eqn:finalul}
\end{equation}
\noindent Using the measured central value for $\mathcal{B}(\btn)$ and taking the known values of
$G_F$, $m_B$, $m_{\tau}$ and $\tau_{B}$ from Ref.~\cite{ref:pdg2004} we calculate, from equation~\ref{eqn:br}, 
the product of the $B$ meson decay constant
and $|\Vub|$ to be
$f_{B}\cdot|\Vub| = (7.0^{+2.3}_{-3.6}(\mbox{stat.})^{+0.4}_{-0.5}(\mbox{syst.}))\times10^{-4}$~GeV.

\clearpage
\newpage
\section{ACKNOWLEDGMENTS}
\label{sec:Acknowledgments}
%
We are grateful for the 
extraordinary contributions of our \pep2\ colleagues in
achieving the excellent luminosity and machine conditions
that have made this work possible.
The success of this project also relies critically on the 
expertise and dedication of the computing organizations that 
support \babar.
The collaborating institutions wish to thank 
SLAC for its support and the kind hospitality extended to them. 
This work is supported by the
US Department of Energy
and National Science Foundation, the
Natural Sciences and Engineering Research Council (Canada),
Institute of High Energy Physics (China), the
Commissariat \`a l'Energie Atomique and
Institut National de Physique Nucl\'eaire et de Physique des Particules
(France), the
Bundesministerium f\"ur Bildung und Forschung and
Deutsche Forschungsgemeinschaft
(Germany), the
Istituto Nazionale di Fisica Nucleare (Italy),
the Foundation for Fundamental Research on Matter (The Netherlands),
the Research Council of Norway, the
Ministry of Science and Technology of the Russian Federation, 
Ministerio de Educaci\'on y Ciencia (Spain), and the
Particle Physics and Astronomy Research Council (United Kingdom). 
Individuals have received support from 
the Marie-Curie IEF program (European Union) and
the A. P. Sloan Foundation.

\end{document}